\documentclass{article}
\usepackage{spconf,amsmath,graphicx}
\usepackage{subfig}
\usepackage{multirow}
\usepackage{amsfonts}
\usepackage{mathrsfs}

\PassOptionsToPackage{hyphens}{url}\usepackage{hyperref}
\title{Domain Generalization with Fourier Transform and Soft Thresholding}
\name{\begin{tabular}{c}Hongyi Pan$^*$, Bin Wang$^*$, Zheyuan Zhang$^*$, Xin Zhu$^\dagger$, Debesh Jha$^*$,\\ Ahmet Enis Cetin$^\dagger$, Concetto Spampinato$^\ddagger$, Ulas Bagci$^*$\end{tabular}
\thanks{This work was supported by the National Institutes of Health (NIH) under grants \#U01 DK127384-02S1, \#R01 CA246704, and \#R01 CA240639.
Code: \url{https://github.com/phy710/ICASSP2024-FDG-ST}.}}
\address{$^*$Machine \& Hybrid Intelligence Lab, Department of Radiology, Northwestern University, USA\\
$^\dagger$Department of Electrical and Computer Engineering, University of Illinois Chicago, USA\\
$^\ddagger$Department of Electrical, Electronic and Computer Engineering, University of Catania, Italy}
\begin{document}
%
\maketitle
\small
\begin{abstract}
Domain generalization aims to train models on multiple source domains so that they can generalize well to unseen target domains. Among many domain generalization methods, Fourier-transform-based domain generalization methods have gained popularity primarily because they exploit the power of Fourier transformation to capture essential patterns and regularities in the data, making the model more robust to domain shifts. The mainstream Fourier-transform-based domain generalization swaps the Fourier amplitude spectrum while preserving the phase spectrum between the source and the target images. However, it neglects background interference in the amplitude spectrum. To overcome this limitation, we introduce a soft-thresholding function in the Fourier domain. 
We apply this newly designed algorithm to retinal fundus image segmentation, which is important for diagnosing ocular diseases but the neural network's performance can degrade across different sources due to domain shifts. The proposed technique basically enhances fundus image augmentation by eliminating small values in the Fourier domain and providing better generalization. The innovative nature of the soft thresholding fused with Fourier-transform-based domain generalization improves neural network models' performance by reducing the target images' background interference significantly. Experiments on public data validate our approach's effectiveness over conventional and state-of-the-art methods with superior segmentation metrics. 
\end{abstract}
\section{Introduction}
Signal processing is the foundation of true intelligence, and the discrete Fourier transform (DFT) algorithm is one of the most important and powerful signal and image processing methods~\cite{brigham1967fast}. In recent years, Fourier-transform-based domain generalization (FDG) methods have been introduced in~\cite{yang2020fda, xu2021fourier}, and Liu~\textit{et al.}~\cite{liu2021feddg} extended the FDG into federated learning. These methods are implemented by swapping and mixing the Fourier amplitude spectrum between the source and the target images and retaining the Fourier phase spectrum. The motivation behind these methods is that the amplitude spectrum captures the lower-level characteristics, such as the image style, while the phase spectrum encapsulates the higher-level semantics, such as the content and objects of the image. However, the amplitude spectrum is influenced by both object and background. Hence, filtering in the Fourier domain to reduce the influence from the background can improve the quality of the augmented images. 

To address this challenge, we introduce a simple yet quite effective approach: a \textit{soft-thresholding} function in the Fourier domain. While the soft-thresholding function is commonly used in wavelet transform domain denoising~\cite{donoho1995noising} and as a proximal operator for the $\ell_1$-norm-based optimization problems~\cite{karakucs2020simulation}, it has never been used in domain generalization setting previously. In this study, we apply it to remove the small entries in the Fourier domain. This procedure is similar to image coding and transform domain denoising~\cite{pan2023real, pan2023hybrid}. To the best of our knowledge, this is the first work that combines frequency-domain denoising with the Fourier-transform-based domain generalization. 

\textbf{Clinical motivation of the application:} Retinal fundus image segmentation plays a pivotal role in the early diagnosis and management of various ocular diseases, such as diabetic retinopathy and glaucoma~\cite{wang2020dofe}. Accurate segmentation of retinal structures, including optic cup and optic disc, is therefore essential for facilitating computer-aided diagnosis and tracking disease progression. However, the performance of automated segmentation methods often degrades when applied to fundus images from different institutions or scanning devices due to domain shifts, variations in image quality, and inconsistent illumination conditions. 

Domain generalization addresses the challenge of building models that can apply their learned knowledge effectively to new and diverse domains~\cite{blanchard2011generalizing, zhou2022domain}. In the initial phases of domain generalization research, the prevailing approach follows the distribution alignment idea in domain adaptation by learning domain invariant features via kernel methods~\cite{motiian2017unified}, meta regularization~\cite{balaji2018metareg}, gradual fine-tuning~\cite{xu2021gradual}, domain adversarial learning~\cite{li2019episodic, zhang2023domain}, and gradient-guided dropout~\cite{hoffman2018cycada}.

Our major contributions are summarized as follows: We introduce the soft-thresholding function into the FDG framework. It improves the fundus image augmentation performance of the neural network model, by reducing the background inference from the target image. The remainder of this paper is organized in this way: In Section~\ref{sec: Methodology}, we state the problem and introduce the traditional FDG method, and then we present how we implement soft thresholding in the FDG. In Section~\ref{sec: Experimental Results}, we introduce the retinal fundus datasets we employ in this work and present our experimental comparison with other state-of-the-art works. In Section~\ref{sec: Conclusion}, we draw our conclusion.

\section{Methodology}~\label{sec: Methodology}
\subsection{Problem Statement}
Medical images are often collected from different clinical centers. Therefore, their distribution usually varies. Consider a set of $K$ training datasets $\{\mathcal{X}_0, \mathcal{X}_1, \dots, \mathcal{X}_{K-1}\}$ with corresponding annotations $\{\mathcal{Y}_0, \mathcal{Y}_1, \dots, \mathcal{Y}_{K-1}\}$ and an test dataset $\{\mathcal{X}_K\}$ from an unseen domain are from different institutions. The goal of this work is to learn a model using $\{\mathcal{X}_0, \mathcal{X}_1, \dots, \mathcal{X}_{K-1}\}$, such that it can directly generalize to a completely unseen testing domain with a high performance. To achieve this goal, we use domain generalization to generate continuous domain distribution using $\{\mathcal{X}_0, \mathcal{X}_1, \dots, \mathcal{X}_{K-1}\}$.

\begin{figure}[tb]
\centering
\subfloat[Domain 1~\cite{sivaswamy2015comprehensive}.]{\includegraphics[page=1, width=0.25\linewidth]{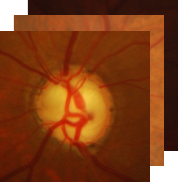}}
\subfloat[Domain 2~\cite{fumero2011rim}.]{\includegraphics[page=2, width=0.25\linewidth]{figures/domain.pdf}} 
\subfloat[Domain 3~\cite{orlando2020refuge}.]{\includegraphics[page=3, width=0.25\linewidth]{figures/domain.pdf}} 
\subfloat[Domain 4~\cite{orlando2020refuge}.]{\includegraphics[page=4, width=0.25\linewidth]{figures/domain.pdf}}
\caption{Samples of ROIs of the fundus images. }
\label{fig: samples of ROIs}
\end{figure}

Samples of the fundus images we use in this work are visualized in Fig.~\ref{fig: samples of ROIs}. They are collected from $K=4$  datasets from diverse domains. More details of the datasets are discussed in Section~\ref{sec: Experimental Results}. 

\subsection{Fourier-Transform-Based Domain Generalization}
A Fourier-transform-based domain generalization framework (FDG) was proposed by Xu~\textit{et al.} in~\cite{xu2021fourier}. It implements data augmentation by swapping the low-frequency magnitude in the Fourier spectrum between the source image and the target image. Specifically, given a source image $\mathbf{x}\in\mathbb{R}^{C\times H\times W}$, its Discrete Fourier Transformation (DFT) $\mathbf{X}=\mathscr{F}(\mathbf{x})$ is formulated as:
\begin{equation}
    \mathbf{X}(c, u, v) = \sum_{h=0}^{H-1}\sum_{w=0}^{W-1} \mathbf{x}(c, h, w)\cdot e^{-j2\pi\left(\frac{hu}{H}+\frac{wv}{W}\right)},
\end{equation}
where $C$ denotes the number of (color) channels, $H$ and $W$ stand for the height and the weight of the image~\cite{brigham1967fast}, respectively. The Fourier representation can be further decomposed to an amplitude spectrum $\mathbf{A}=\text{abs}(\mathbf{X})$ and a phase spectrum $\mathbf{P}=\angle{\mathbf{X}}$. The amplitude spectrum reflects the low-level distribution (\textit{e.g.} style), and the phase spectrum reflects the high-level semantics (\textit{e.g.} object) of the image. Therefore, we can revise the style of an image by changing its amplitude and retaining its phase. The data argumentation to transfer the style of the image to the target with the amplitude of $\tilde{\mathbf{A}}$ can be implemented as:
\begin{equation}
    \hat{\mathbf{A}} = (1-\lambda)\mathbf{A}+\lambda\tilde{\mathbf{A}},\label{eq: fda}
\end{equation}
where $\lambda\in(0, 1]$ controls the strength of the augmentation. If $\lambda=1$, the amplitude spectrum is just swapped. 

Then, the mixed amplitude spectrum $\hat{\mathbf{A}}$ is combined with the original phase spectrum $\mathbf{P}$ to construct the augmented Fourier representation:
\begin{equation}
    \hat{\mathbf{X}}(c, u, v) =  \hat{\mathbf{A}}(c, u, v)\cdot e^{-j\mathbf{P}(c, u, v)}.
\end{equation}

Finally, the augmented image $\hat{\mathbf{x}}=\mathscr{F}^{-1}(\hat{\mathbf{X}})$ is obtained by Inverse Discrete Fourier Transformation (IDFT):
\begin{equation}
    \hat{\mathbf{x}}(c, h, w) = \frac{1}{HW}\sum_{u=0}^{H-1}\sum_{v=0}^{W-1} \hat{\mathbf{X}}(c, u, v)\cdot e^{j2\pi\left(\frac{hu}{H}+\frac{wv}{W}\right)}.
\end{equation}

\begin{figure}[tb]
\centering
\subfloat[Domain generalization on the ROI of the Image. ]{\includegraphics[page=1, width=1\linewidth]{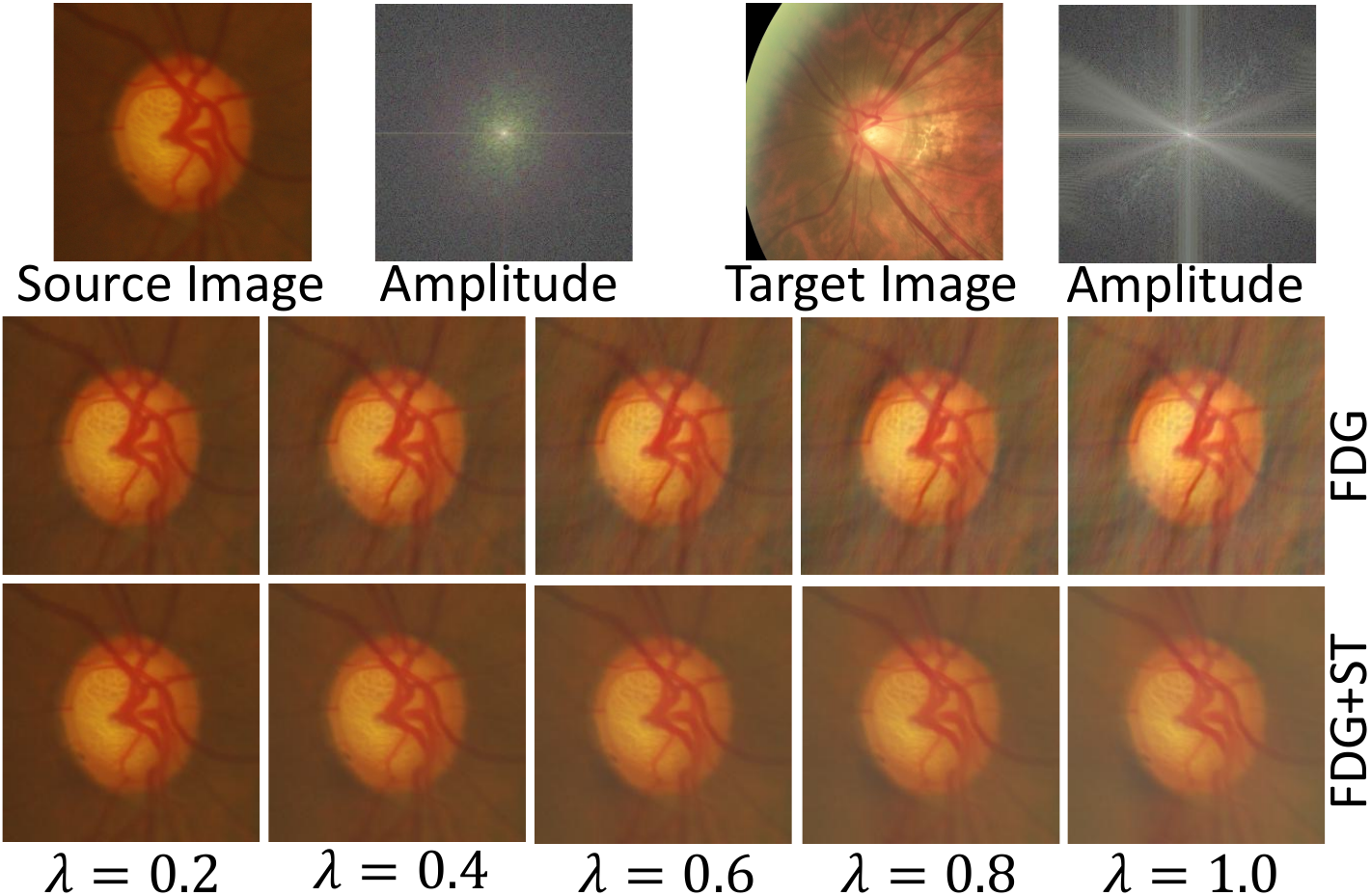}}\\
\subfloat[Domain generalization on the whole image. ]{\includegraphics[page=2, width=1\linewidth]{figures/demo1.pdf}}
\caption{Samples of the Fourier-transform-based domain generalization. In this work, domain generalization is applied to the ROI of the images that are provided directly by~\cite{wang2020dofe}, and it can also be extended to the whole images.}
\label{fig: demo}
\end{figure}

Fig.~\ref{fig: demo} demonstrates augmented samples from a traditional FDG approach~\cite{xu2021fourier}. The amplitude spectrums are visualized by taking the logarithm and then being normalized to $[0, 1]$.  As $\lambda$ increases, the style of the source images changes to the style of the target image. However, the inference from the background cannot be neglected, especially when $\lambda=1.0$. In Fig.~\ref{fig: demo}, especially in the most right sub-figures, some aliasing (shadows) artifacts can be observed in the augmented results.

\subsection{Soft-Thresholding-Improved Fourier-Based Domain Generalization}
\begin{figure}[tb]
\centering
\subfloat[$\alpha$ = 0.1\%.]{\includegraphics[width=0.19\linewidth]{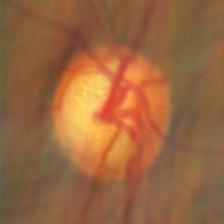}}\vspace{1pt}
\subfloat[$\alpha$ = 1\%.]{\includegraphics[width=0.19\linewidth]{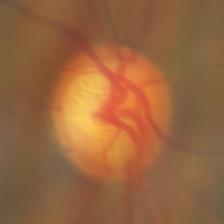}}\vspace{1pt}
\subfloat[$\alpha$ = 5\%.]{\includegraphics[width=0.19\linewidth]{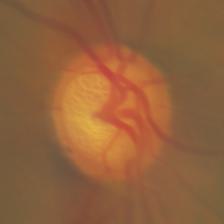}}\vspace{1pt}
\subfloat[$\alpha$ = 10\%.]{\includegraphics[width=0.19\linewidth]{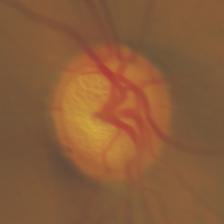}}\vspace{1pt}
\subfloat[$\alpha$ = 50\%.]{\includegraphics[width=0.19\linewidth]{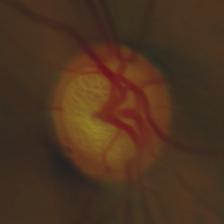}}
\caption{Grid search on $\alpha$. For the best visualization on augmentation, we set $\lambda=1.0$ here.}
\label{fig: alpha}
\end{figure}

\begin{figure}[htbp]
\centering
\includegraphics[width=1\linewidth]{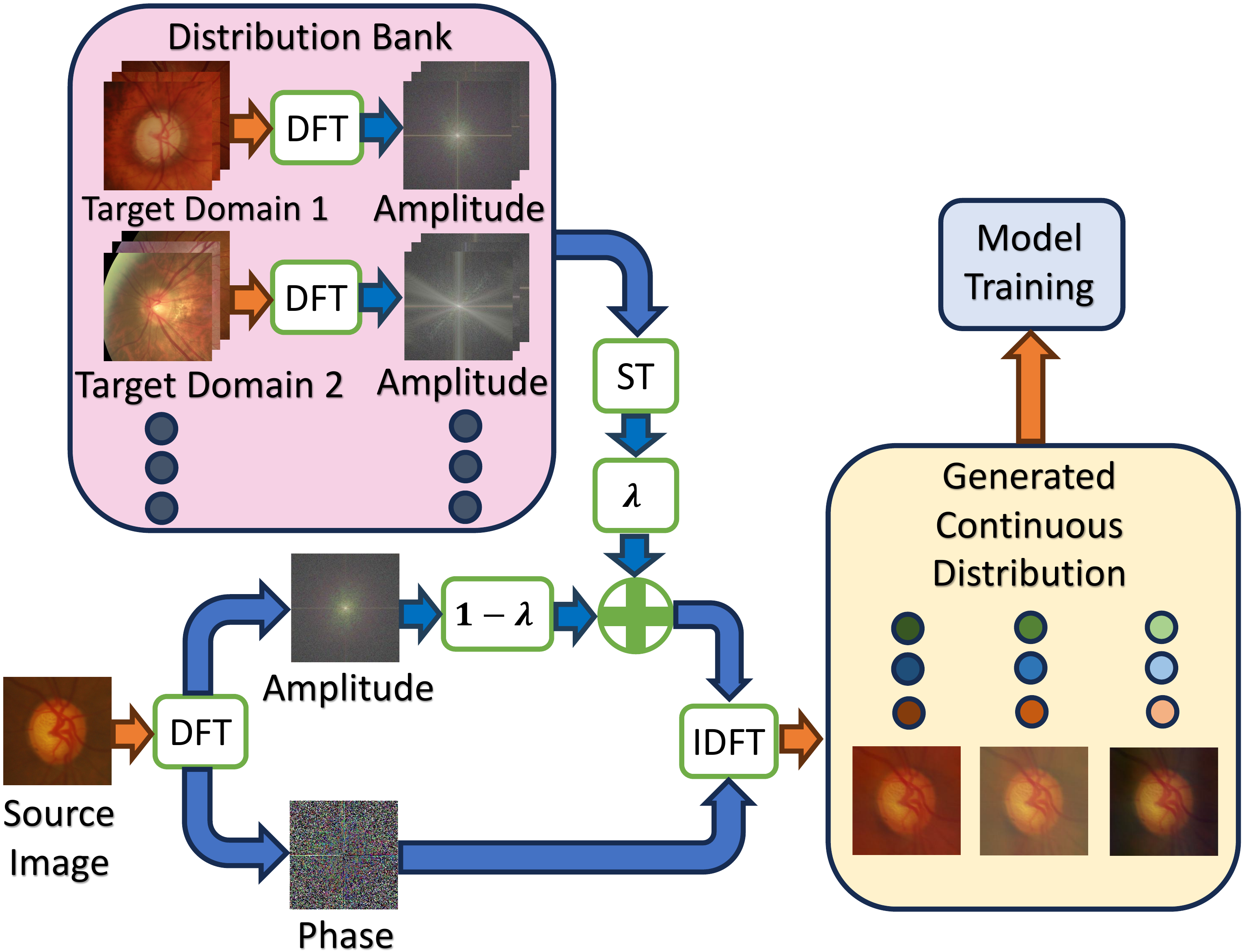}
\caption{The proposed FDG framework with ST.}
\label{fig: framework}
\end{figure}

Soft thresholding (ST) is a mathematical operation commonly used in signal processing, statistics, and data analysis. It is often employed in denoising or compressing data, especially when dealing with wavelet transforms or other sparse signal representations. The primary purpose of soft thresholding is to reduce the magnitude of certain values in a signal while preserving the significant features and minimizing the impact of noise. Furthermore, soft thresholding can be used to reduce the background in certain data analysis and signal processing applications. It's a technique commonly employed to separate meaningful signal components from unwanted noise or background interference~\cite{donoho1995noising, tomita2022denoising}. The soft-thresholding function is defined as: 
\begin{equation}
    \mathcal{S}(X, T) = \text{sign}(X)\cdot \max\{|X|-T, 0\},
\end{equation}
where $X$ is the input feature and $T$ is the threshold parameter. In this work, we reduce the background inference of the target image by the soft-thresholding function. Instead of mixing the source amplitude spectrum with the target amplitude spectrum, we fuse the source amplitude with the soft-thresholding-filtered target amplitude spectrum. In detail, we revise Eq. (\ref{eq: fda}) as:
\begin{equation}\label{eq: fdast}
    \hat{\mathbf{A}} = (1-\lambda)\mathbf{A}+\lambda\mathcal{S}(\hat{\mathbf{A}}, \mathbf{T}),
\end{equation}
where $\mathbf{T}\in\mathbb{R}^C$ are the soft threshold parameters, and $C=3$ in this task. Therefore, specifically, we use 3 individual thresholds for the red, green, and blue channels respectively. Each threshold is determined by the DFT amplitude in the corresponding channel. Herein, we set the soft threshold parameters are set as $5\%$ of the largest magnitudes, \textit{i.e.}, 
\begin{equation}\label{eq: T=amaxA}
   \mathbf{T}=\alpha\cdot\max(\hat{\mathbf{A}}), 
\end{equation}
where $\alpha=5\%$, determined by the grid search, and demonstrated in Fig.~\ref{fig: alpha}. Note that the optimal $\alpha$ is not unique because the augmented images are similar when $\alpha$ increases or decreases ``slightly''. However, a too-large $\alpha$ can make the image darker. This is because if the threshold values are too large, many amplitude entries will be filtered to 0s. On the other hand, the soft-thresholding function with a too-small $\alpha$ cannot reduce the inference from the target background. As Fig.~\ref{fig: demo} shows, the soft-thresholding function improves the FDG augmented results by eliminating the aliasing (shadows) artifacts. The overall framework of FDG with ST is demonstrated in Fig.~\ref{fig: framework}.

\begin{table*}[tb]
\centering
\begin{tabular}{cc|cccc|c|cccc|c}
\hline\noalign{\smallskip}
\multirow{3}{*}{\bf{Method}}&\multirow{3}{*}{\bf{Metric}}&\multicolumn{5}{c|}{\bf{Optic Cup Segmentation}}&\multicolumn{5}{c}{\bf{Optic Disc Segmentation}}\\
& & \multicolumn{4}{c|}{\bf{Unseen Domain}}&\multirow{2}{*}{\bf{Average}}& \multicolumn{4}{c|}{\bf{Unseen Domain}}&\multirow{2}{*}{\bf{Average}}\\
& & 1 & 2&  3  &4 & & 1 & 2&  3  &4 &\\
\noalign{\smallskip}\hline\noalign{\smallskip}
\multirow{3}{*}{DoFE}&DSC$\uparrow$&80.86&77.97&83.95&\bf{87.67}&82.61&\bf{95.11}&89.05&91.25&\bf{93.50}&92.23\\
&HD$\downarrow$&37.47&29.15&21.42&\bf{14.50}&25.64&17.17&31.10&21.63&14.86&21.19\\
&ASD$\downarrow$&19.67&15.75&11.17&7.05&13.41&\bf{8.51}&16.96&12.29&\bf{7.16}&11.23\\
\noalign{\smallskip}\hline\noalign{\smallskip}
\multirow{3}{*}{FDG}&DSC$\uparrow$&80.75&\bf{79.27}&83.62&87.56&82.80&95.08&88.71&92.01&93.39&92.29\\
&HD$\downarrow$&36.80&\bf{27.10}&21.64&14.77&25.08&\bf{16.71}&29.25&20.61&\bf{14.50}&20.27\\
&ASD$\downarrow$&19.61&14.25&11.33&\bf{7.02}&13.05&8.52&17.19&11.23&7.29&11.06\\
\noalign{\smallskip}\hline\noalign{\smallskip}
\multirow{3}{*}{\bf{FDG+ST}}&DSC$\uparrow$&\bf{81.64}&78.93&\bf{84.11}&86.83&\bf{82.88}&94.72&\bf{89.20}&\bf{92.51}&92.91&\bf{92.34}\\
&HD$\downarrow$&\bf{35.36}&27.22&\bf{21.29}&15.91&\bf{24.95}&17.56&\bf{27.40}&\bf{19.77}&15.05&\bf{19.95}\\
&ASD$\downarrow$&\bf{18.80}&\bf{13.91}&\bf{11.04}&7.66&\bf{12.85}&9.12&\bf{16.05}&\bf{10.54}&7.79&\bf{10.88}\\
\noalign{\smallskip}\hline\noalign{\smallskip}
\end{tabular}
\caption{Segmentation experiment results. }
\label{tab: experimental results}
\end{table*}

\section{Experimental Results}~\label{sec: Experimental Results}
In this work, we follow publicly available retinal fundus image optic cup and optic disc segmentation challenge tasks in~\cite{wang2020dofe}. It consists of 4 different clinical centers: Domain 1 is the Drishti-GS dataset~\cite{sivaswamy2015comprehensive}, domain 2 is RIM-ONE-r3 dataset~\cite{fumero2011rim}, domain 3 and domain 4 are from REFUGE train and validation subsets~\cite{orlando2020refuge}, respectively. These datasets were collected using different scanners, patient profiles, and demographics. More details of these datasets can be found in Table I in~\cite{wang2020dofe}. For each image, we use a region of interest (ROI) with a size of $800 \times 800$. These ROIs are provided directly by~\cite{wang2020dofe} via a sample U-Net~\cite{ronneberger2015u}, and they are visualized in Fig.~\ref{fig: samples of ROIs}. 

We compare our proposed method with a strong baseline: the domain-oriented feature embedding (DoFE)~\cite{wang2020dofe} approach. While this method adopts the \textit{Domain Knowledge Pool} to learn and memorize the prior knowledge from multi-source domains, FDG-based approaches~\cite{yang2020fda, xu2021fourier, liu2021feddg} augment the images directly. For a fair comparison, all methods employ DeepLab-V3+~\cite{chen2018encoder} with MobileNet-V2~\cite{sandler2018mobilenetv2} as a backbone for the basic segmentation framework, and the ROI images are down-sampled to $256\times 256$ for less computational cost. We follow the common practice in domain generalization literature to adopt the leave-one-domain-out strategy for the evaluation, \textit{i.e.}, we train the model on 3 datasets and use the remaining one (the unseen domain) for testing. For each method, we train 4 models for the 4 unseen domains. Each model accomplishes the optic cup segmentation task and the optic disc segmentation task simultaneously. To train the models with DoFE, we first pre-train the vanilla DeepLabV3+ network for 40 epochs with a learning rate of 0.001 without domain generalization. Next, we apply the domain generalization methods on the training dataset and continue to train the whole framework for another 80 epochs with a learning rate of 0.001 and then for 60 epochs with a learning rate of 0.0002. In FDG, we set $\lambda$ in Eq.~(\ref{eq: fda}) and Eq.~(\ref{eq: fdast}) dynamically in the range of (0.0, 1.0], and in ST, we set $\alpha=5\%$ in Eq.~(\ref{eq: T=amaxA}). 
We train the models with FDG and the models with FDG and ST using the same number of epochs and the same learning rates as how we train the models with DoFE. We employ the Adam optimizer~\cite{kingma2014adam} and set the batch size as 16. The experiments are carried out on a workstation computer with an NVIDIA RTX3090 GPU using the PyTorch library~\cite{paszke2019pytorch}. We evaluate methods based on their Hausdorff distance (HD), average surface distance (ASD)~\cite{bilic2023liver}, and dice similarity coefficient (DSC) on the unseen domain test dataset: 
\begin{equation}
    d(\hat{y}, Y)=\inf_{\hat{y}\in \hat{Y}} ||y-\hat{y}||,
\end{equation}
\begin{equation}
    HD(Y, \hat{Y}) = \max\bigg\{\sup_{y\in Y} d(y, \hat{Y}), \sup_{\hat{y}\in \hat{Y}} d(\hat{y}, Y)\bigg\},
\end{equation}
\begin{equation}
    ASD(Y, \hat{Y}) = \frac{1}{|Y|+|\hat{Y}|}\left(\sum_{y\in Y}d(y, \hat{Y})+ \sum_{\hat{y}\in \hat{Y}}d(\hat{y}, Y)\right),
\end{equation}
\begin{equation}
    DSC(Y, \hat{Y})=\frac{2TP(Y, \hat{Y})}{2TP(Y, \hat{Y})+FP(Y, \hat{Y})+FN(Y, \hat{Y})},
\end{equation}
where $Y$ is the segmentation ground-truth, $\hat{Y}$ is the segmentation result, $||\cdot||$ denotes the Euclidean distance, and $|\cdot|$ indicates the size of the region. $TP$, $FP$, and $FN$ stand for the numbers of pixel-level true positives, true negatives, false positives, and false negatives, respectively.
Therefore, lower HD and ASD as well as higher DSC indicate a better model performance.

\begin{figure}[tb]
\centering
    \subfloat{\includegraphics[width=0.23\linewidth]{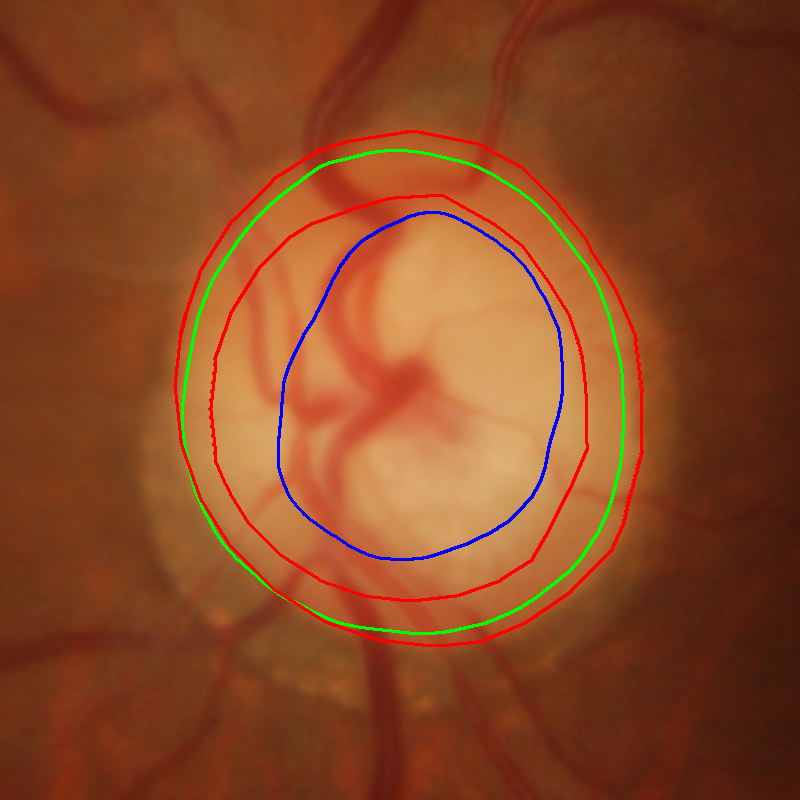}}\vspace{1pt}
\subfloat{\includegraphics[width=0.23\linewidth]{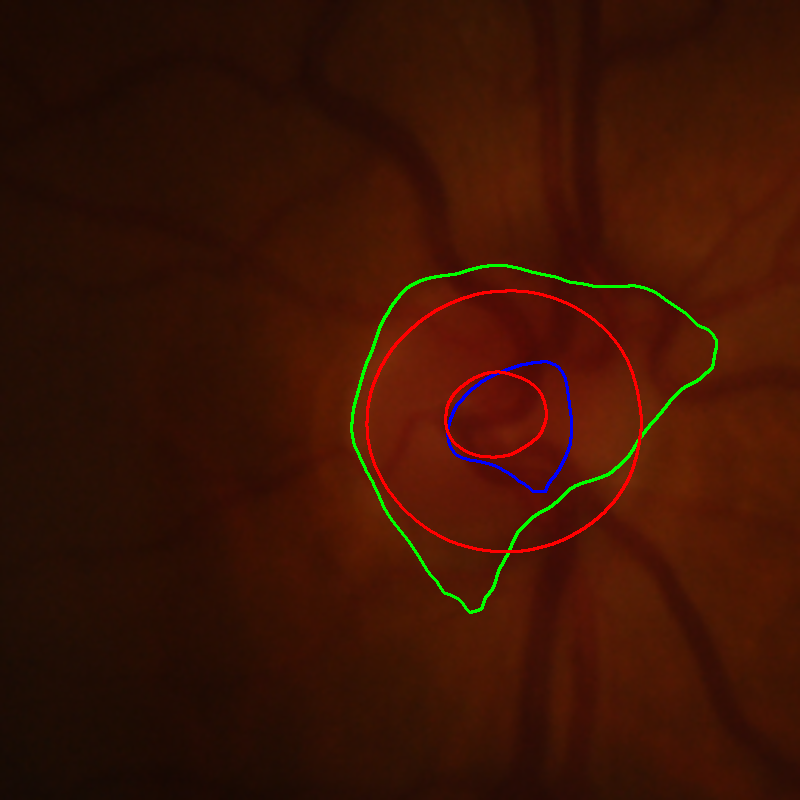}}\vspace{1pt}
\subfloat{\includegraphics[width=0.23\linewidth]{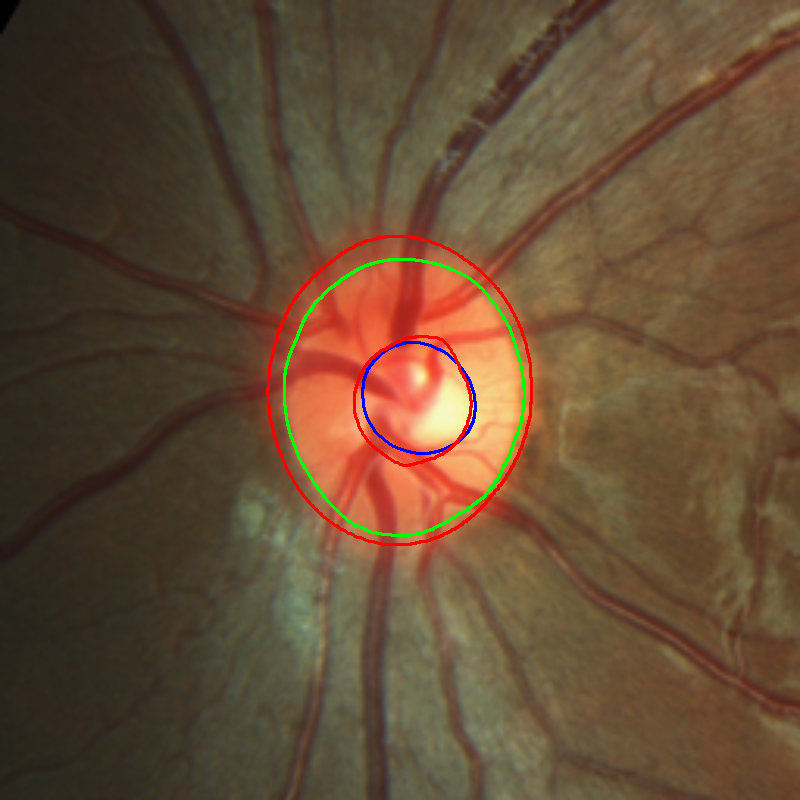}}\vspace{1pt}
\subfloat{\includegraphics[width=0.23\linewidth]{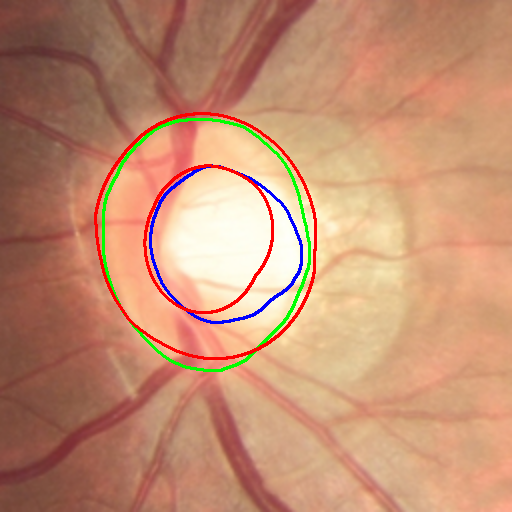}}\\\hspace{-5pt}
\subfloat{\includegraphics[width=0.23\linewidth]{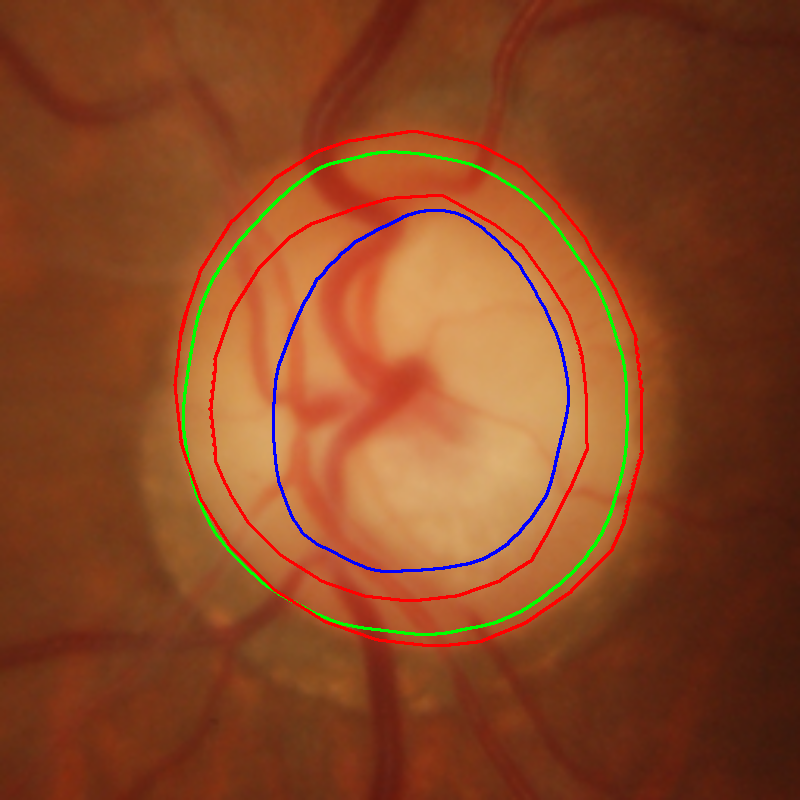}}\vspace{1pt}
\subfloat{\includegraphics[width=0.23\linewidth]{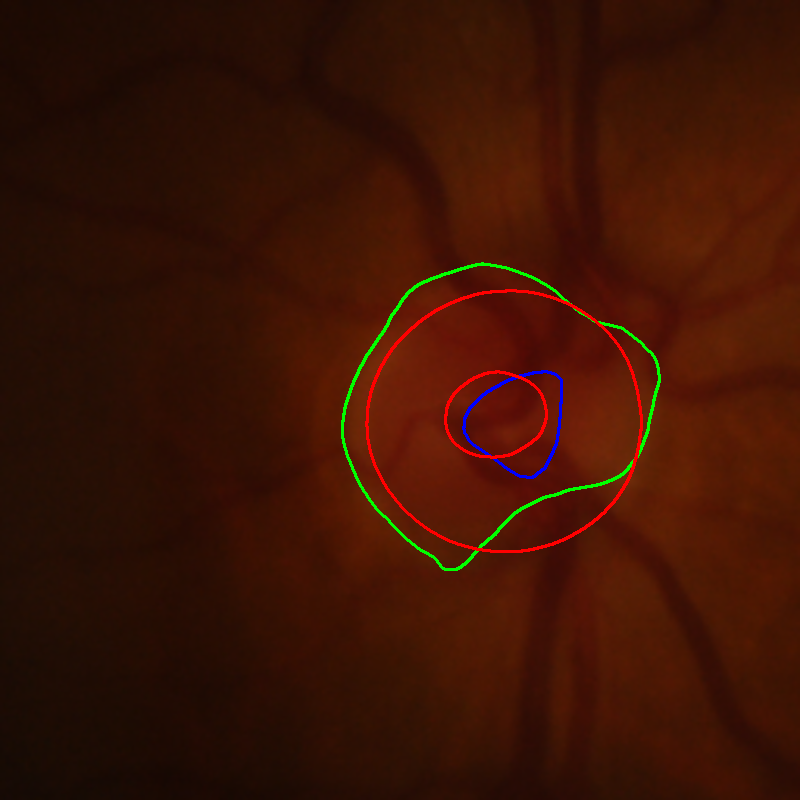}}\vspace{1pt}
\subfloat{\includegraphics[width=0.23\linewidth]{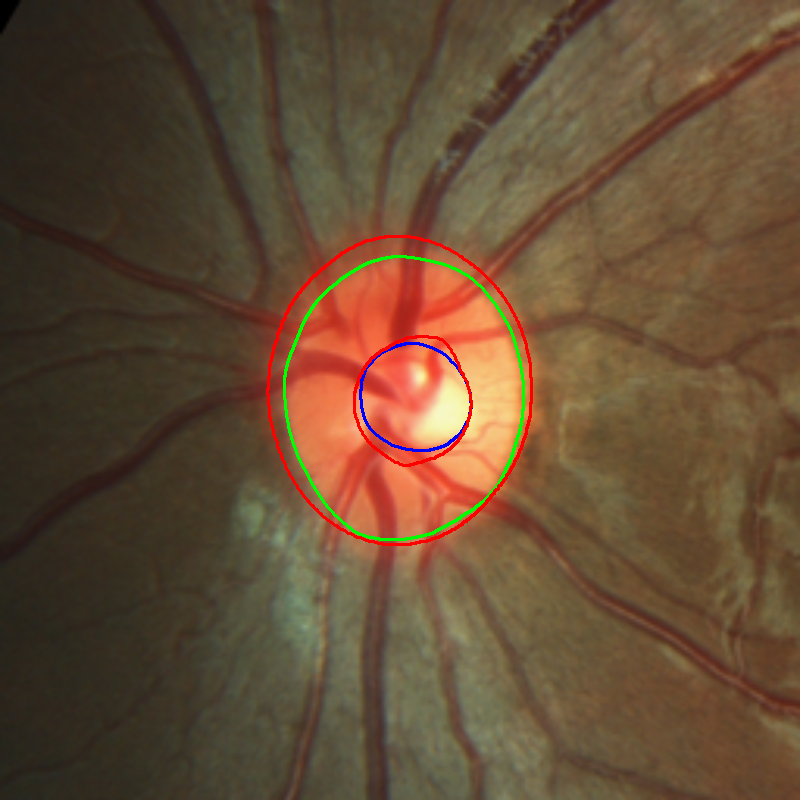}}\vspace{1pt}
\subfloat{\includegraphics[width=0.23\linewidth]{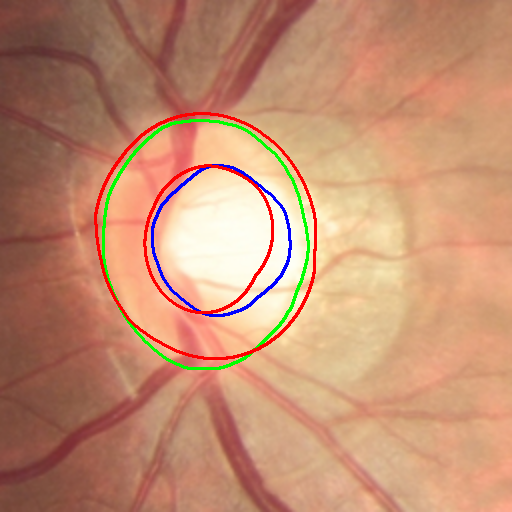}}\\\hspace{-5pt}
\subfloat{\includegraphics[width=0.23\linewidth]{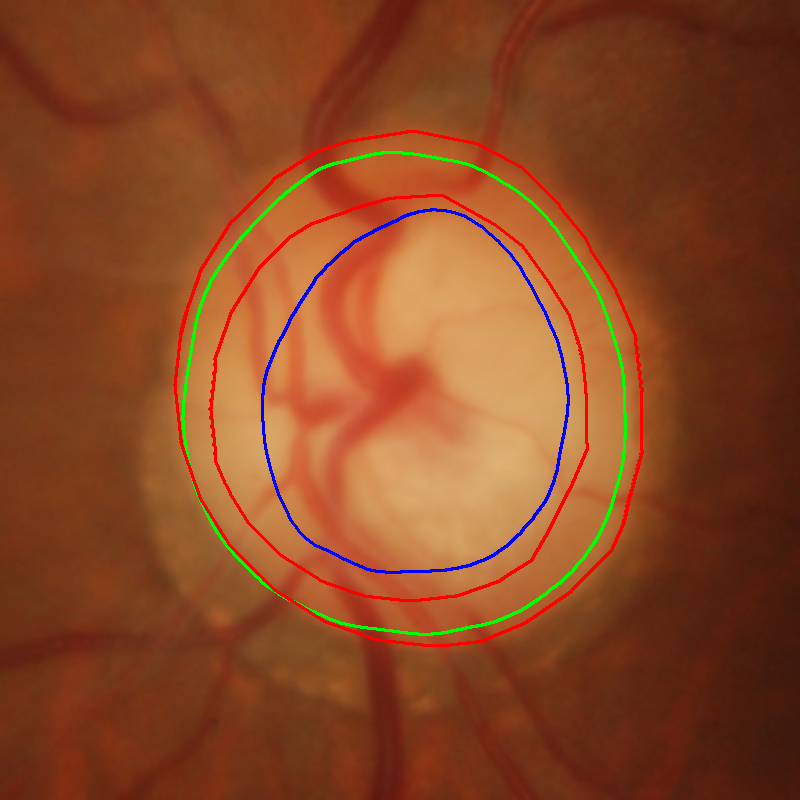}}\vspace{1pt}
\subfloat{\includegraphics[width=0.23\linewidth]{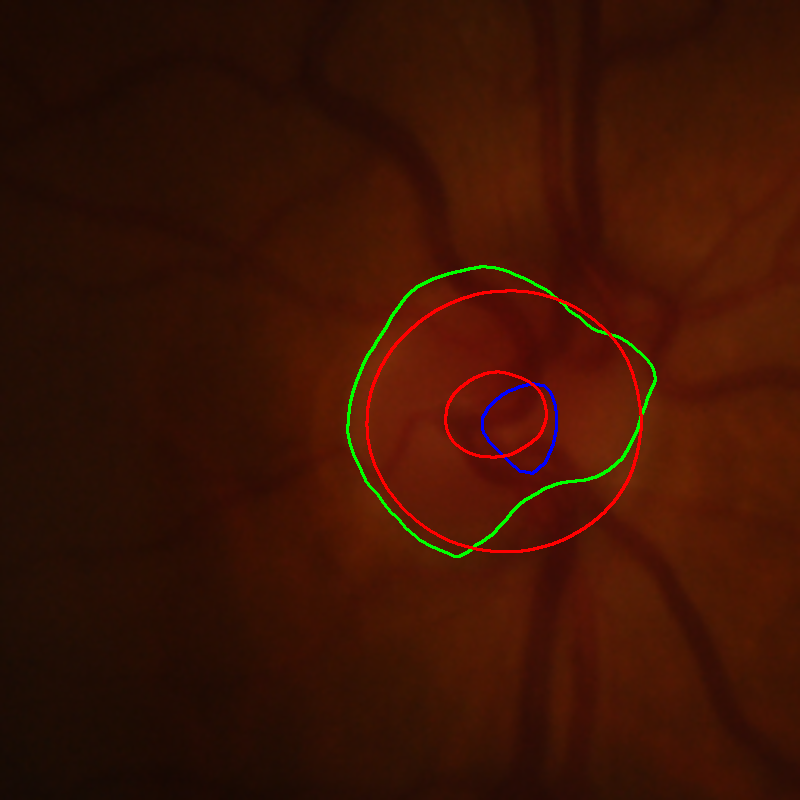}}\vspace{1pt}
\subfloat{\includegraphics[width=0.23\linewidth]{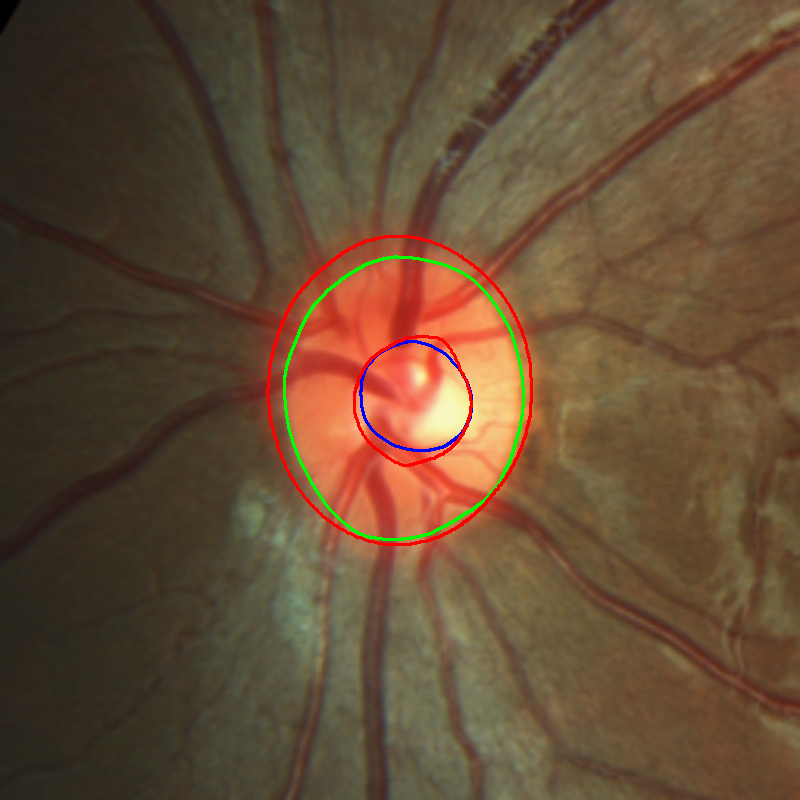}}\vspace{1pt}
\subfloat{\includegraphics[width=0.23\linewidth]{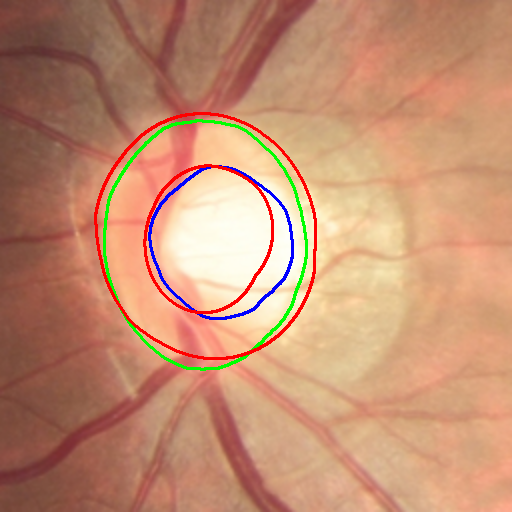}}\\
\caption{Samples of the results. From top to bottom: DoFE, FDG, FDG+ST. From left to right: Domain 1/2/3/4. The red, blue, and green contours present the boundaries of the ground truths, the optic cups, and the optic discs, respectively. }
\label{fig: result}
\end{figure}

The quantitative comparison with recent domain generalization methods on the OC/OD segmentation tasks is presented in Table~\ref{tab: experimental results}. On average, introducing the soft-thresholding function into the FDG improves the performance of the model significantly. Example results with improved FDG samples are demonstrated in Fig.~\ref{fig: demo}, and samples of the segmentation results are show in Fig.~\ref{fig: result}.

\section{Conclusion}\label{sec: Conclusion}
In this paper, we showed that Fourier-Transform-based domain generalization (FDG) methods are preferable over the conventional domain generalization methods. In particular, we have demonstrated a simple yet effective strategy to improve FDG-based strategies by introducing a soft-thresholding function. While the traditional FDG methods swap the Fourier amplitude spectrum between the source and the target images and contain the phase spectrum, they neglect background interference in the amplitude spectrum. As a result, the aliasing and shadows can be observed in the augmented images if the parameters are not set properly. The proposed integration of soft-thresholding function in the FDG setting separates meaningful signal components from unwanted noise or background interference significantly better than FDG-based approaches. We showed that our direct and simple approach can increase the average dice similarity coefficient and decrease the average Hausdorff distance and surface distance of the model. 

\small
\bibliographystyle{IEEEbib}
\bibliography{strings,refs}

\end{document}